\documentclass[preprints,article,accept,moreauthors,pdftex]{Definitions/mdpi}
\firstpage{1} 
\pubvolume{1}
\issuenum{1}
\articlenumber{0}
\pubyear{2021}
\copyrightyear{2020}
\datereceived{} 
\dateaccepted{} 
\datepublished{} 
\hreflink{https://doi.org/} % If needed use \linebreak
\pdfoutput=1

\DeclareMathOperator{\Tr}{Tr}

\Title{Laser Cooling Beyond Rate Equations: Approaches From Quantum Thermodynamics}

\TitleCitation{Laser Cooling Beyond Rate Equations: Approaches from Quantum Thermodynamics}

\Author{C. N. Murphy $^{1}$\orcidA{}, L. Toledo Tude $^{1}$\orcidB{} and P. R. Eastham $^{1}$*\orcidC{}}

\AuthorNames{Firstname Lastname, Firstname Lastname and Firstname Lastname}

\AuthorCitation{Murphy, C. N.; Toledo Tude, L.; Eastham, P. R.}
\address[1]{%
$^{1}$ \quad School of Physics, Trinity College Dublin, Ireland}

\corres{Correspondence: easthamp@tcd.ie}

\abstract{Solids can be cooled by driving impurity ions with lasers,
  allowing them to transfer heat from the lattice phonons to the
  electromagnetic surroundings. This exemplifies a quantum thermal
  machine, which uses a quantum system as a working medium to transfer
  heat between reservoirs. We review the derivation of the
  Bloch-Redfield equation for a quantum system coupled to a reservoir,
  and its extension, using counting fields, to calculate heat
  currents. We use the full form of this equation, which makes only
  the weak-coupling and Markovian approximations, to calculate the
  cooling power for a simple model of laser cooling. We compare its
  predictions with two other time-local master equations: the secular
  approximation to the full Bloch-Redfield equation, and the Lindblad
  form expected for phonon transitions in the absence of driving. We
  conclude that the full Bloch-Redfield equation provides accurate
  results for the heat current in both the weak- and strong- driving
  regimes, whereas the other forms have more limited
  applicability. Our results support the use of Bloch-Redfield
  equations in quantum thermal machines, in spite of their potential
  to give unphysical results.}


\begin{document}
%%%%%%%%%%%%%%%%%%%%%%%%%%%%%%%%%%%%%%%%%%
%\setcounter{section}{-1} %% Remove this when starting to work on the template.

\section{Introduction}

Laser cooling
\cite{seletskiy_laser_2016,nemova_laser_2010,epstein_optical_2009}, in
both atomic and solid-state systems, is now a well established
technique. In solids, particularly rare-earth doped glasses, cooling
can be achieved by using anti-Stokes fluorescence of the dopants. It
provides an example of a quantum thermal machine
\cite{scovil_three-level_1959,geusic_quantum_1967,linden_how_2010}, in
which a discrete quantum system -- in this case, the energy levels of
a rare earth ion -- is the working medium. This working medium couples
to two heat baths and a source of work, namely the phonon and photon
reservoirs and the driving laser, allowing it to operate as a
refrigerator.

Laser cooling is generally modelled using rate equations for the
populations of the levels. This approach can also be used for
semiconductors, where the rate equations refer to the populations of
the electron and hole bands. However, such approaches cannot capture
certain effects which, while not expected to be relevant in systems
such as rare-earths, are increasingly important in quantum
thermodynamics more generally. These include the role of coherences in
determining heat flows, which have been argued to offer enhanced
performances in various quantum thermal machines
\cite{creatore_efficient_2013,fruchtman_photocell_2016,dorfman_efficiency_2018,kilgour_coherence_2018,murphy_quantum_2019};
the effects of strong driving, which can modify the energy levels
through the a.c. Stark effect\
\cite{brash_dynamic_2016,eastham_lindblad_2013,ramsay_damping_2010},
and so impact on the heat flows\
\cite{murphy_quantum_2019,gauger_heat_2010}; and the effects of
spectral structure in the heat baths. This last can be considered in
two regimes: for strongly structured baths one can expect
non-Markovian behaviour\
\cite{abiuso_non-markov_2019,thomas_thermodynamics_2018,bylicka_thermodynamic_2016},
whose impact on thermodynamics remains a challenging open
topic. However, spectral structure can be important even where a
Markovian description remains appropriate\
\cite{correa_quantum-enhanced_2015}. An important practical target for
thermodynamic machines is to maximize their power, and the heat flows
to a bath are determined by its spectral density. Thus to achieve
maximum power one must consider the spectral structure of baths, if
there is any on the energy scales of the working medium. Examples of
systems where this occurs include quantum-dot excitons coupled to
acoustic phonons\ \cite{nazir_modelling_2016}, colour centres in
diamond\ \cite{wrachtrup_processing_2006,norambuena_microscopic_2016},
and superconducting circuits\ \cite{basilewitsch_reservoir_2019}.

These issues can be treated theoretically by studying models of an
open quantum system in which the working medium interacts with its
surrounding heat baths. Such models are tractable in the
weak-coupling, Markovian regime, where they lead to time-local
equations of motion such as the Bloch-Redfield equation\
\cite{breuer_theory_2002}. Those approaches can be extended to allow
calculations of heat and work in the quantum regime\
\cite{esposito_nonequilibrium_2009}. However, there are several
time-local equations which can be obtained, using reasonable
approximations, from a given model, and these can make differing
predictions for the dynamics
\cite{eastham_bath-induced_2016,hofer_markovian_2017}. This problem
has been addressed by several groups, who argue that the Bloch-Redfield
equation\
\cite{eastham_bath-induced_2016,hartmann_accuracy_2020,purkayastha_out--equilibrium_2016,kilgour_coherence_2018,liu_coherences_2021,jeske_bloch-redfield_2015,kilgour_path-integral_2019,boudjada_dissipative_2014}
is useful and indeed accurate, despite its potential pathologies\
\cite{dumcke_proper_1979}. In this paper we extend such studies to
explore the heat flows in a simple laser cooling process, with the aim
of identifying an approximate time-local equation that can accurately
model them.  

%validity of time-local master equations for other thermal machines are 
%thermal machines, includining the quantum absorption refrigerator,

In the following, we first review the derivation of the Bloch-Redfield
equation for an open quantum system, and outline its extension to
calculate heat flows. We also discuss two other time-local equations
which can be obtained on making further approximations: a Lindblad
form in the energy eigenbasis, obtained by making the secular
approximation, and a Lindblad form in the eigenbasis of the undriven
system. We use these forms to calculate the cooling spectrum, i.e. the
cooling power as a function of driving frequency, in a model of laser
cooling. The model allows for strong driving and includes a spectral
structure for the environment. We find that a complete description of
the cooling spectrum, which covers both the weak-driving and
strong-driving regimes, can be achieved using the full Bloch-Redfield
equation. We provide further support for the correctness of the
Bloch-Redfield master equation -- whose use has been controversial
because it does not guarantee positivity\ \cite{dumcke_proper_1979},
and can lead to behaviour inconsistent with thermodynamic principles\
\cite{gonzalez_testing_2017} -- by comparing its predictions to those
of an exact numerical method. Our conclusions support the use of
Bloch-Redfield equations to model laser cooling and other
thermodynamic processes\
\cite{kilgour_coherence_2018,liu_coherences_2021,friedman_quantum_2018,trushechkin_open_2021,kilgour_path-integral_2019,boudjada_dissipative_2014}.

%The introduction should briefly place the study in a broad context and highlight why it is important. It should define the purpose of the work and its significance. The current state of the research field should be reviewed carefully and key publications cited. Please highlight controversial and diverging hypotheses when necessary. Finally, briefly mention the main aim of the work and highlight the principal conclusions. As far as possible, please keep the introduction comprehensible to scientists outside your particular field of research.
%
%Citing a journal paper \cite{ref-journal}. Now citing a book reference \cite{ref-book1,ref-book2} or other reference types \cite{ref-unpublish,ref-communication,ref-proceeding}. Please use the command \citep{ref-thesis,ref-url} for the following MDPI journals, which use author--date citation: Administrative Sciences, Arts, Econometrics, Economies, Genealogy, Histories, Humanities, IJFS, Journal of Intelligence, Journalism and Media, JRFM, Languages, Laws, Religions, Risks, Social Sciences.
 
%%%%%%%%%%%%%%%%%%%%%%%%%%%%%%%%%%%%%%%%%%
\section{Materials and Methods}

\subsection{Laser Cooling Model}

\begin{figure}[H]
\begin{center}\includegraphics{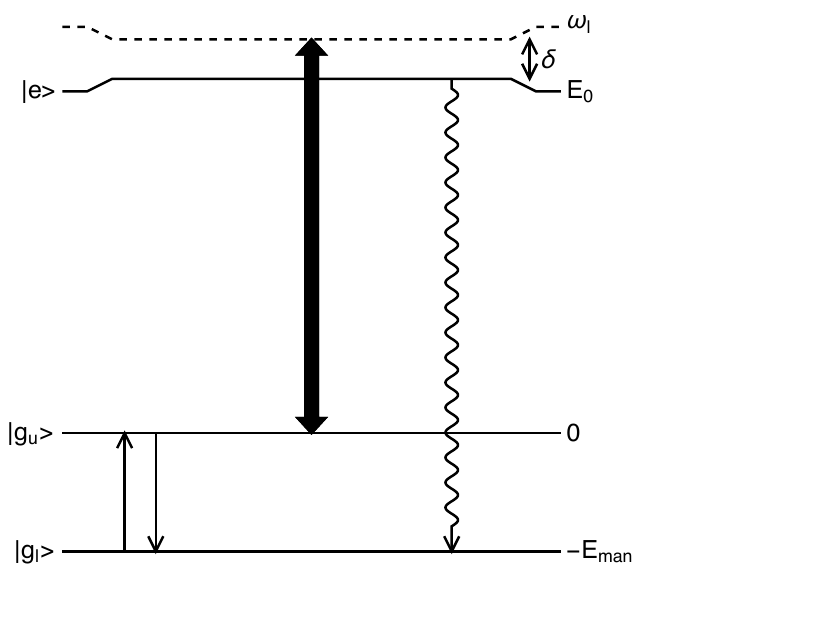}\end{center}
\caption{Energy levels of an impurity in a model laser-cooling process. The two states of a ground-state manifold, $|g_u\rangle$ and $|g_l\rangle$, are coupled by the emission and absorption of lattice phonons (vertical solid lines). Laser driving occurs on the transition from the upper level of the ground-state manifold to an excited state $|e\rangle$ (block arrow). This state decays radiatively to the ground state (wavy arrow). \label{figschema}}
\end{figure}   

We consider a simple laser-cooling scheme, depicted in Fig.\ \ref{figschema}, involving an impurity with two states forming a ground-state manifold, and a single state in an excited-state manifold. The two states within the ground-state manifold, $|g_l\rangle$, $|g_u\rangle$, are split by an energy $E_{\mathrm{man}}$, and coupled by the emission and absorption of lattice phonons. The driving laser excites the transition from the upper state of the ground-state manifold to an excited state $|e\rangle$ an energy $E_0$ above. We assume this state decays by radiative emission to the ground-state. We use $\Omega$ to denote the Rabi splitting of the driven transition,  and $\delta=\omega_l-E_0$ the driving laser frequency relative to the transition.

The time-dependence of the electric field driving the transition can be removed, in the rotating wave approximation, by using a unitary transformation \begin{equation} U=\exp(i\omega_l t |e\rangle\langle e|).\end{equation} In this frame the field of the laser is time-independent, and the Hamiltonian for the system is \begin{equation} H_S=\begin{pmatrix} -\delta & \Omega/2 & 0 \\ \Omega/2 & 0 & 0 \\ 0 & 0 & -E_{\mathrm{man}}\end{pmatrix}.\label{eqsysham}\end{equation} The diagonal terms here are the energies of the electronic states, in the rotating frame. The off-diagonal terms are the coupling between those states produced by the electric-dipole interaction with the driving field\ \cite{allen_optical_1987}. Note that here, and throughout this paper, we set $\hbar=1$.

\subsection{Master equations for open quantum systems}

Fig.\ \ref{figschema} depicts an open quantum system: one which interacts, explicitly or implicitly, with a wider environment. These interactions lead to an exchange of energy between system and environment, and dephasing and decoherence effects. Here, we have an environment comprising the phonons in the host crystal of the impurity, and the photons associated with the radiative decay of the upper level. 

The dynamics of an open quantum system can, in certain circumstances,
be described by a time-local master equation for its reduced density
matrix\ \cite{breuer_theory_2002}. Such equations can be obtained from
microscopic models which consider the environment explicitly on making
the weak-coupling and Markovian approximations. They are also often
postulated phenomenologically, based on the observation that the most
general equation-of-motion is one of Lindblad form. However, there are
several different forms of equations which can result from a
microscopic model, depending on the details of the approximations
made. The predictions of these forms can, furthermore, differ from
those based on phenomenological Lindblad forms.

These issues have been discussed in previous works\
\cite{eastham_bath-induced_2016,hartmann_accuracy_2020} which suggest
that the full Bloch-Redfield equation -- obtained by using the
weak-coupling and Markovian approximations, but without making the
secular approximation -- gives a good description of the
dynamics. This is in spite of the fact that the Bloch-Redfield
equation does not guarantee that the eigenvalues of the reduced
density matrix remain positive\
\cite{dumcke_proper_1979,breuer_theory_2002}. For a system where there
are no degeneracies, or near-degeneracies, that issue can be cured by
secularization\ \cite{dumcke_proper_1979,wangsness_dynamical_1953},
which corresponds to eliminating oscillating terms in the dissipator
that average to zero over time. This leads to a Lindblad
form~\cite{lindblad_generators_1976,gorini_completely_1976} with
positive rates. It is, however, a priori invalid for the laser-cooling
protocol considered here, where weak driving near to resonance means
we have $\Omega\approx 0$ and $\delta\approx 0$, so that two of the
eigenstates of Eq. (\ref{eqsysham}), $|g_u\rangle$ and $|e\rangle$,
are almost degenerate in the frame where the laser field is
time-independent.

A fairly generic form for the Hamiltonian of an open quantum system
is \begin{align}
H&=H_{S}+H_{B}+H_{SB} \label{eqopen}\\
H_{SB}&=\sum_k g_k O (b_k+b_k^\dagger).\label{eqcoupling}\end{align} 
Here $H_{S}$ is the Hamiltonian for the system, $H_{B}$ for its environment, or bath, and $H_{SB}$ is the system-bath coupling. We consider the common situation in which the bath comprises a set of harmonic oscillators\ \cite{breuer_theory_2002}, which we index using a quantity or quantities labelled $r$. Note that $r$ denotes the full set of quantum numbers required to label the modes. The oscillators have 
frequencies $\omega_r$, and ladder operators $b_r$ and $b_r^\dagger$. The displacement of the $r^{\mathrm{th}}$ bath mode is coupled to the system operator $O$, with coupling strength $g_r$. The dissipative effects of the bath depend on its spectral density, $J(\omega)=\sum_r g_r^2\delta(\omega-\omega_r)$. 

To fix notation we recall the standard procedure for deriving a
Bloch-Redfield master
equation~\cite{breuer_theory_2002,redfield_theory_1957,bloch_generalized_1957}. We
work in the interaction picture with respect to $H_{S}+H_{B}$, so that
$O(t)=e^{iH_S t} O e^{-iH_S t}$. Note that where necessary we will
distinguish operators in the interaction and Schr\"odinger pictures
as, for example, $O(t)$ and $O$. Iterating the von Neumann equation
gives the form \begin{equation}\frac{d\rho(t)}{dt}=-\int^t dt^\prime
  [H_{SB}(t),[H_{SB}(t^\prime),\rho(t^\prime)]],\label{eqpertdyn}\end{equation}
where $\rho(t)$ is the full density operator of the system and
environment. For weak coupling to a bath one can replace
$\rho(t^\prime)\approx \rho_S(t^\prime)\otimes \rho_B(t^\prime)$ on
the right-hand side, where $\rho_S$ is the reduced density matrix of
the system, and $\rho_B$ that of the bath. Since the bath is
macroscopic it can be assumed to be unperturbed by the system, and
$\rho_B$ taken to be a thermal state at inverse temperature $\beta$.
For a Markovian system one may, furthermore, approximate
$\rho_S(t^\prime)\approx \rho_S(t)$. We can write the coupling
operator in the eigenbasis of $H_{S}$ as \begin{equation}
  O(t)=\sum_{ij} e^{i(E_i-E_j)t} \langle i|O|j\rangle |i\rangle\langle
  j| \equiv \sum_{ij} \hat{O}_{ij}(t).\end{equation} Taking the trace
of Eq. (\ref{eqpertdyn}) over the environment's degrees-of-freedom we
find\begin{equation}\begin{split} \frac{d\rho_S(t)}{dt}= \sum_{ij}
    \big\{ A_{ij} & [\hat{O}_{ji}(t)\rho_S(t)O(t)+
    O(t)\rho_S(t)\hat{O}_{ij}(t) \\ & \qquad\qquad\qquad
    -\rho_S(t)\hat{O}_{ij}(t)O(t)-O(t)\hat{O}_{ji}(t)\rho_S(t) ] \\
    -iB_{ij} & [\hat{O}_{ji}(t)\rho_S(t)O(t)-
    O(t)\rho_S(t)\hat{O}_{ij}(t) \\ & \qquad\qquad\qquad
    +\rho_S(t)\hat{O}_{ij}(t)O(t)-O(t)\hat{O}_{ji}(t)\rho_S(t)
    ]\big\}.\end{split}\label{eqnonsecmaster}\end{equation}

The quantities $A_{ij}$ and $B_{ij}$ are related to the the real-time Green's functions of the environment at the transition frequency $\nu_{ij}=E_i-E_j$ connecting levels $i$ and $j$. The quantities $A_{ij}$ are associated with dissipation, and are \begin{equation} A_{ij}=\pi \{[n(\nu_{ij})+1]J(\nu_{ij})+n(\nu_{ji})J(\nu_{ji})\}.\label{eqrates}\end{equation}Here $n(\nu>0)=1/(\exp(\beta \nu)-1)$ is the Bose function describing the bath occupation, and $J(\nu)=0$ for $\nu<0$. The first term in $A_{ij}$ corresponds to the creation of a bath quantum as the system transitions from a state $i$ to $j$ with $E_i-E_j>0$, whereas the second corresponds to the absorption of a bath quantum in the opposite case, $E_i-E_j<0$. The quantities $B_{ij}$ are associated with energy shifts, and are given by the principal value integral \begin{equation} B_{ij}=\mathcal{P} \int J(\omega) \frac{\omega +(2n(\omega)+1)(E_i-E_j)}{\omega^2-(E_i-E_j)^2}d\omega.\end{equation}

Eq. (\ref{eqnonsecmaster}) can be used directly, but is often further
approximated, leading to other forms of equation-of-motion for an open
quantum system. One very common approximation is to drop the
principal value terms proportional to $B_{ij}$. Another common
approximation is to \emph{secularize} the equation-of-motion. This is
done by decomposing the remaining coupling operators, $O(t)$, into the
energy eigenbasis: $O(t)=\sum_{kl}\hat{O}_{kl}(t)$. Every term in
Eq. (\ref{eqnonsecmaster}) then involves a product of operators
corresponding to two transitions, one involving the pair of levels $i$ and
$j$, and one involving the pair $k$, $l$. If the levels are non-degenerate
these products of operators are, in general, time-dependent in the
interaction picture, and average to zero. The exception is where a
transition in one direction is paired with the same transition in the
opposite direction, so that the time-dependence cancels out. Retaining
only those terms the dissipative part of Eq. (\ref{eqnonsecmaster}) becomes \begin{equation}\begin{split} \frac{d\rho_S(t)}{dt}= \sum_{ij}
   2A_{ij}\left(\hat{O}_{ji}(t)\rho_S(t)\hat{O}_{ij}(t)-\frac{1}{2}[\rho_S(t),\hat{O}_{ij}(t)\hat{O}_{ji}(t)]_+\right),\end{split}\label{eq:secmaster}\end{equation} where $[A,B]_+=AB+BA$ is an anticommutator. This is of Lindblad form, and therefore guarantees the positivity of the density operator. It has a straightforward physical interpretation: the environment causes transitions from the system state $i$ to the system state $j$ at rate $2A_{ij}$. 

\subsection{Heat flows from master equations}

The method of full counting statistics
\cite{esposito_nonequilibrium_2009} allows one to extend the
approaches above so as to compute the heat transferred to the bath. It
has been used, often with the secular approximation
\cite{silaev_lindblad-equation_2014}, to obtain master equations and
study heat statistics in various systems, including driven quantum-dot
excitons \cite{murphy_quantum_2019,gauger_heat_2010}, a driven
two-level system \cite{gasparinetti_heat-exchange_2014}, a
steady-state (absorption) refrigerator
\cite{friedman_quantum_2018,liu_coherences_2021,kilgour_coherence_2018,friedman_quantum_2018},
and a two-bath spin-boson model
\cite{kilgour_path-integral_2019,boudjada_dissipative_2014}. The
absorption refrigerator and spin-boson model have been studied using
the full Bloch-Redfield approach, without the secular approximation,
which highlights the role of coherences\
\cite{liu_coherences_2021,kilgour_coherence_2018,friedman_quantum_2018}.
Here we give an outline of the method and present a complete form for
the full counting-field Bloch-Redfield equation, which we shall use to
calculate laser cooling spectra.

The heat transferred to a bath is, by definition, the change in its
energy between two times. Thus we consider a process involving
projective measurements of the bath energy at two times. We take the
initial time to be $t_i=0$, and suppose that at this time the system
and bath are in a product state, $\rho_S(0)\otimes\rho_B$. We can then
consider the probability distribution of the heat, $P(Q,t)$, which is
the probability that the energy measurements of the bath at times
$t_i$ and $t$ give results differing by $Q$. It is convenient also to
introduce the characteristic function of the heat distribution,
$\chi(u,t)=\int dQ P(Q,t) e^{iuQ}$. The variable $u$ is known as the
counting field. (This term should not be taken to imply that heat is
necessarily a discrete, countable quantity. It arises from other uses
of the method, such as calculations of the number of electrons
transferred across a tunnel junction\
\cite{esposito_nonequilibrium_2009}.)

One can evaluate $\chi(u,t)$ by introducing an annotated density
operator, $\rho_u(t)$, such that $\chi(u,t)=\Tr
\rho_u(t)$. $\rho_u(t)$ has a non-unitary time evolution given by
\begin{equation}\rho_u(t)=U_{u/2}\rho_u(0)U_{-u/2}^\dagger,\label{eqnonunit}\end{equation} where $U_{u}$ is related
to the normal time-evolution operator, $U=e^{-iHt}$, by \begin{equation}
  U_u=e^{iuH_B}Ue^{-iuH_B}.\label{eqphasemarker}\end{equation} Note
the similarity between these phase factors and the factor $e^{iuQ}$ in
the definition of the characteristic function; it is these factors
that incorporate the results of the measurements of the bath energy,
$H_B$, into $\rho_u(t)$. At the initial time the annotated density matrix is given by $\rho_u(0)=\rho(0)$. 

For a general operator $P$ we define the annotated version
$P_u=e^{iuH_B}Pe^{-iuH_B}$, which obeys the Heisenberg-like
equation \begin{displaymath}
  i\frac{dP_u}{du}=[P_u,H_B]. \end{displaymath} For the lowering
operator appearing in Eq. (\ref{eqcoupling}) we have
$b_{u,k}=e^{-i\omega_k u}b_{0,k}$. Thus the time-evolution operators,
$U_{\pm u/2}$, can be obtained from the standard form, $e^{-iHt}$, by
replacing the coupling Hamiltonian, Eq. (\ref{eqcoupling}), with
$H^{\pm}_{SB}=\sum g_k O (b_k e^{\mp i\omega u/2}+b_k^\dagger e^{\pm
  i\omega u/2})$.

A master equation for the reduced annotated density matrix,
$\rho_{u,S}(t)$ can now be obtained, following the steps above. The
essential difference is that the von Neumann equation for $\rho(t)$,
in the interaction picture, must be replaced
by \begin{equation}\frac{d\rho_u(t)}{dt}=-i
  (H_{SB}^+\rho_u(t)-\rho_u(t) H_{SB}^-).\end{equation} The result
is \begin{equation}\begin{split} \frac{d\rho_{u,S}(t)}{dt}= \sum_{ij}
    \big\{ A_{ij} & [e^{iu(E_i-E_j)}(\hat{O}_{ji}(t)\rho_{u,S}(t)O(t)+
    O(t)\rho_{u,S}(t)\hat{O}_{ij}(t)) \\ & \qquad
    -\rho_{u,S}(t)\hat{O}_{ij}(t)O(t)-O(t)\hat{O}_{ji}(t)\rho(t) ] \\
    -iB_{ij} & [e^{iu(E_i-E_j)}(\hat{O}_{ji}(t)\rho_{u,S}(t)O(t)-
    O(t)\rho_{u,S}(t)\hat{O}_{ij}(t)) \\ & \qquad
    +\rho_{u,S}(t)\hat{O}_{ij}(t)O(t)-O(t)\hat{O}_{ji}(t)\rho(t)
    ]\big\}.\end{split}\label{eqnonseccountf}\end{equation} This form differs
from Eq. (\ref{eqnonsecmaster}) by the addition of phase factors in
the four terms that cause transitions between the system
eigenstates. It can approximated as discussed above, by dropping the
principal value terms, or by making the secular approximation.

The mean heat is \begin{equation}\langle Q\rangle=\int Q P(Q)dQ=-i\left.\frac{d\chi}{du}\right|_{u=0}=-i\Tr \left.\frac{d\rho_{u,S}(t)}{du}\right|_{u=0}.\label{eqmeanheat}\end{equation} From Eq. (\ref{eqnonseccountf}) we find that the heat current is \begin{equation} \begin{split}
    \frac{d\langle Q\rangle}{dt}=\sum_{ij}
    \big\{ A_{ij} & [(E_i-E_j)\Tr (\hat{O}_{ji}(t)\rho_{S}(t)O(t)+
    O(t)\rho_{S}(t)\hat{O}_{ij}(t))] \\
    - iB_{ij} & [(E_i-E_j)\Tr (\hat{O}_{ji}(t)\rho_{S}(t)O(t)-
    O(t)\rho_{S}(t)\hat{O}_{ij}(t))]\big\}.\end{split}\end{equation} This can be used to calculate the heat current from the density matrix, $\rho_{u=0,S}(t)=\rho_{S}(t)$, obtained by solving the standard Bloch-Redfield Eq. (\ref{eqnonsecmaster}). 

\subsection{Master equations for laser cooling}

We consider a model in which the system Hamiltonian is given by
Eq. (\ref{eqsysham}). We suppose that there is a continuum of phonons
responsible for transitions between the states of the ground-state
manifold. This phonon bath will be described by Eqs. (\ref{eqopen})
and (\ref{eqcoupling}), with coupling operator
$O=|g_l\rangle\langle g_u|+|g_u\rangle\langle g_l|$. For the spectral
density of this bath we take the super-Ohmic form with an exponential
high-frequency cut-off,
$J(\omega)=2\alpha(\omega^3/\omega_c^2)\exp(-\omega/\omega_c)$. We are
not targetting a detailed model of a real system, and this form is
chosen largely for illustrative purposes. It may, however, be noted
that it corresponds to that for acoustic phonons coupling to localized
impurities such as the silicon-vacancy centre in diamond
\cite{norambuena_microscopic_2016} or a quantum-dot exciton
\cite{nazir_modelling_2016}. $\alpha$ is a dimensionless measure of
the coupling strength, and $\omega_c$ a high-frequency cut-off. Such
cut-offs arise from the size of the electronic states, and correspond
roughly to the phonon frequency at a wavelength given by that size.

We also consider, in the following, an alternative form of dissipator,
of standard Lindblad form. For a transition caused by a jump operator
$A$, with rate $\gamma_A$, the standard Lindblad form is \begin{equation}
  \frac{d\rho_S(t)}{dt}=\gamma_A\mathcal{L}_A\rho_S(t)=\gamma_A
  \left(A\rho_S(t) A^\dagger -
  \frac{1}{2}[\rho_S(t),A^\dagger A]_+\right).\label{eqlindblad}\end{equation}
Thus the natural phenomenological form, capturing the processes shown
in Fig.\ \ref{figschema}, is to combine two of these dissipative terms, one for
phonon absorption, with rate $\gamma_+$ and jump operator
$\sigma_+=|g_u\rangle\langle g_l|$, and one for phonon emission, with
rate $\gamma_-$ and jump operator
$\sigma_-=\sigma_+^\dagger=|g_l\rangle\langle g_u|$. Such a form
corresponds to Eq.~(\ref{eq:secmaster}) when the eigenstates of $H_S$ are simply
$|g_l\rangle$ and $|g_u\rangle$, which is resonable for weak
driving. This comparison allows us to identify the appropriate rates,
from Eq. (\ref{eqrates}), as $\gamma_-=2\pi (n(E_{\mathrm{man}})+1)J(E_{\mathrm{man}})$ and
$\gamma_+=2\pi n(E_{\mathrm{man}})J(E_{\mathrm{man}})$.

In addition to the phonon dissipation, our model involves
the radiative decay of the excited state, $|e\rangle$ to the ground
state $|g_l\rangle$. We model this as a Lindblad form with jump
operator $|g_l\rangle\langle e|$, and rate $\gamma$.

\subsection{Exact methods}

As well as results of master equations, we shall present, in the
following, calculations of the heat flows obtained by
numerically-exact simulations
\cite{popovic_quantum_2021,strathearn_efficient_2018,fux_efficient_2021}
of the model open quantum system described above. The technique, known
as TEMPO, calculates the path-integral for the evolution of an open
quantum system, discretizing time into a series of steps
\cite{makri_tensor_1995}. It uses a matrix-product state
representation to efficiently store the augmented density tensor,
which allows it to consider large memory times for the bath
\cite{strathearn_efficient_2018,fux_efficient_2021}. Combining path-integral methods with the counting-field
technique \cite{kilgour_path-integral_2019,popovic_quantum_2021},
allows calculations of the total heat transferred to the phonon bath
up to a particular time. Details of the method, and the associated
code, are given in Ref.\ \cite{popovic_quantum_2021}.  We use it to
calculate the heat currents to the phonons by taking the difference of
the total heat transferred to the bath between two times, separated by
a single timestep. In these calculations the dynamics of the system
and effects of the phonon bath are treated exactly. We do not treat
the radiative decay in this first-principles fashion, but rather
include it using the same Lindblad form we use for the master equation
approach. We believe this is appropriate inasmuch as the bath
associated with radiative decay has no spectral structure, in contrast
with that associated with the phonons. The TEMPO approach has recently
been extended to simulations with multiple baths\
\cite{gribben_exact_2021}, which would allow it to treat laser cooling
with structured photon environments, e.g., in optical resonators. 

\section{Results}

The parameters in our model are the energy splitting of the
ground-state manifold, the detuning and Rabi frequency of the driving,
the radiative decay rate, $\gamma$, the cut-off frequency, $\omega_c$,
the dimensionless coupling, $\alpha$, and the temperature $T$. We
choose energy and time units such that $E_{\mathrm{man}}=2$. For the
remaining parameters we take $\gamma=0.5$, $\omega_c=1$,
$\alpha=0.01$, and $T=3$. These parameters are not intended to be
realistic but are chosen so as to allow us to compute the exact
solutions with a reasonable effort, and compare the results of the
different master equations. In particular, we choose a large value for
the radiative decay rate, $\gamma$, to increase the magnitude of the
heat current. It may be noted that for these parameters the phonon
absorption rate, $\gamma_+\approx 0.14$, is comparable to, but smaller
than, the radiative decay rate. This differs from the situation for
conventional laser cooling, appropriate in systems such as rare-earth
ions, where the phonon rates are much larger than those for radiative
decay \cite{seletskiy_laser_2016}, and the electronic populations are
very close to equilibrium. It implies that, in our case, the heat
current will be limited by the driving strength (for weak driving) or
the phonon rate (for strong driving), and not the radiative lifetime.

\begin{figure}[H]
\begin{center}\includegraphics{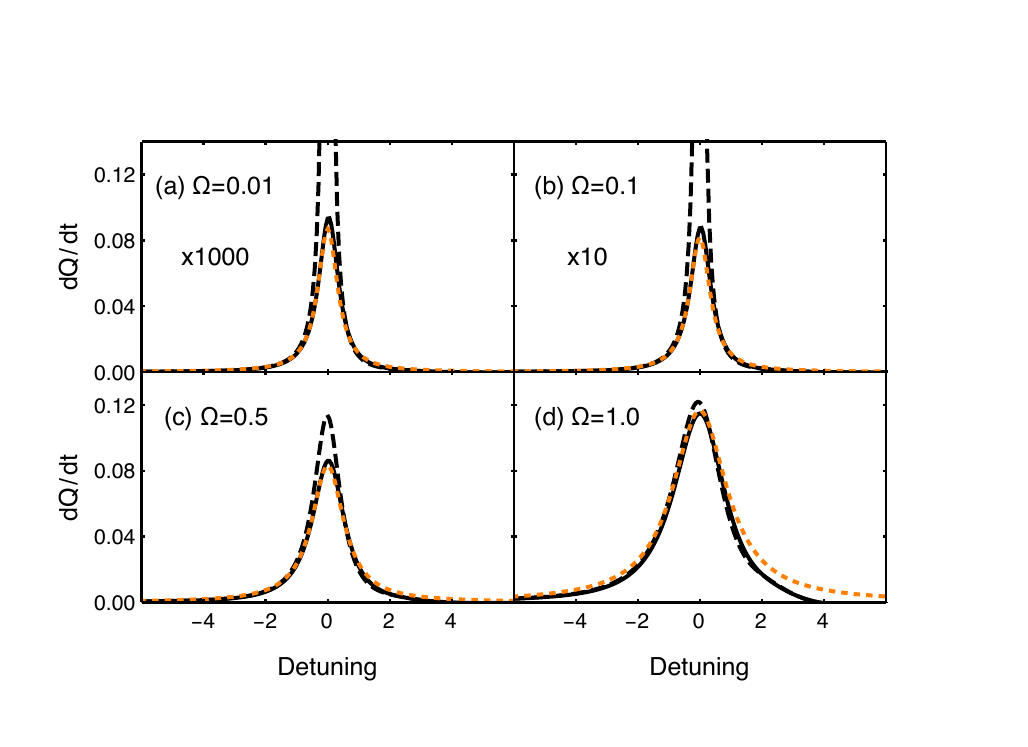}\end{center}
\caption{Rates of heat absorption from the phonon bath, as a function
  of the detuning $\delta=\omega_l-E_0$ of the driving laser from
  resonance, for four different Rabi frequencies. For each Rabi
  frequency we show results computed using the full Bloch-Redfield equation
  (solid black curve), a phenomenological Lindblad equation (dashed
  orange curve), and the Bloch-Redfield equation in the secular
  approximation without the principal value terms (dashed black
  curve). The Rabi frequencies $\Omega$ are: (\textbf{a}) 0.01,
  (\textbf{b}) 0.1, (\textbf{c}) 0.5, and (\textbf{d})
  1.0. \label{figcools1}}
\end{figure}

Fig.\ \ref{figcools1} shows the calculating cooling power as a
function of the detuning, $\delta$, for four different strengths of
driving field. The different curves are computed using the full
Bloch-Redfield equation,~(\ref{eqnonsecmaster}), the phenomenological
Lindblad form, Eq.~(\ref{eqlindblad}), and the secular Bloch-Redfield
equation,~(\ref{eq:secmaster}). Considering first weak driving, in
Fig.\ \ref{figcools1}a, we see that the Bloch-Redfield and phenomenological
theories agree well, and give a cooling profile which appears to be
Lorentzian, as one would expect. While the secular Bloch-Redfield equation
agrees away from the resonance, we see that it fails close to it,
massively overestimating the cooling power. The secular approximation
is, of course, not justified here, because there are near degeneracies
in the Hamiltonian. Nonetheless, the level of disagreement seems
surprising, given the agreement away from resonance.

In the converse, strong-driving region, Fig.\ \ref{figcools1}d, all
three methods give similar results. However, there is a noticeable
difference on the high-energy side of the transition, with the
phemenological theory giving, as before, a Lorentzian profile, while
the other theories predict the heat current drops off more rapidly,
and indeed switches direction, from cooling to heating, in the range
of detunings shown.
\begin{figure}[h]
\begin{center}\includegraphics{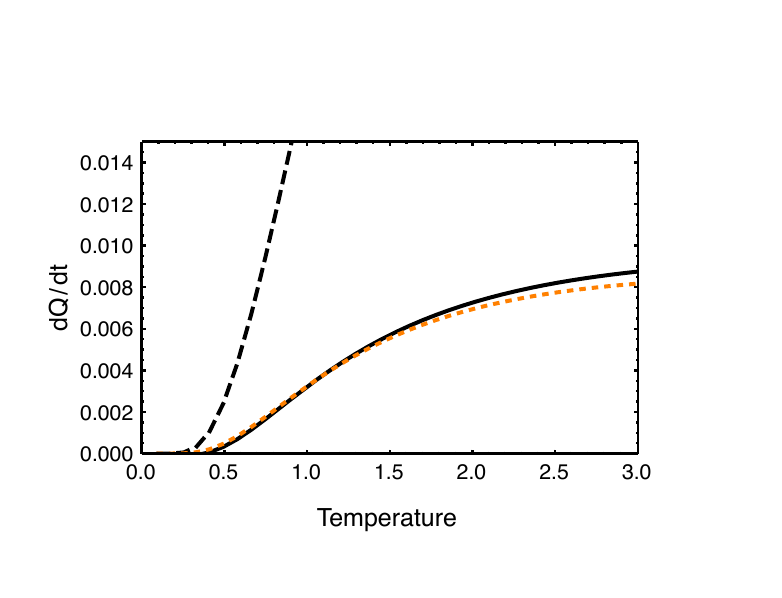}\end{center}
\caption{Rate of heat absorption from the phonon bath, as a function
  of temperature, for the different approaches. Results are shown for weak driving, $\Omega=0.1$, at resonance, $\delta=0$. The heat current is computed using the full Bloch-Redfield equation
  (solid black curve), the phenomenological Lindblad equation (dashed
  orange curve), and the Bloch-Redfield equation in the secular
  approximation without the principal value terms (dashed black
  curve).\label{figtempdep}}
\end{figure}   

Fig.\ \ref{figtempdep} shows the temperature dependence of the cooling power for weak resonant driving. As noted above, the secular approximation is inappropriate in this regime and massively overestimates the cooling power. The other theories agree closely and appear physically reasonable, predicting that the cooling power drops rapidly once the temperature is lowered below the splitting $E_{\mathrm{man}}$. This is the expected physical behaviour for laser cooling in a discrete level structure, caused by the vanishing of the phonon occupation at temperatures much less than $E_{\mathrm{man}}$. The precise behaviour at very low temperatures is not relevant to laser cooling, since it is not expected to operate there. Nonetheless it may be noted that, for these parameters, the Bloch-Redfield theory predicts a very small but negative cooling power, i.e. net heating, at very low temperatures ($T<0.38$). However, this is an approximate theory whose accuracy is not sufficient to discern the true behaviour of the heat current in this regime. Indeed, we find that at these very low temperatures the theory does not predict a physical density matrix, giving one which has a negative, albeit very small, eigenvalue. 

\begin{figure}[H]
\begin{center}\includegraphics[width=10.5 cm]{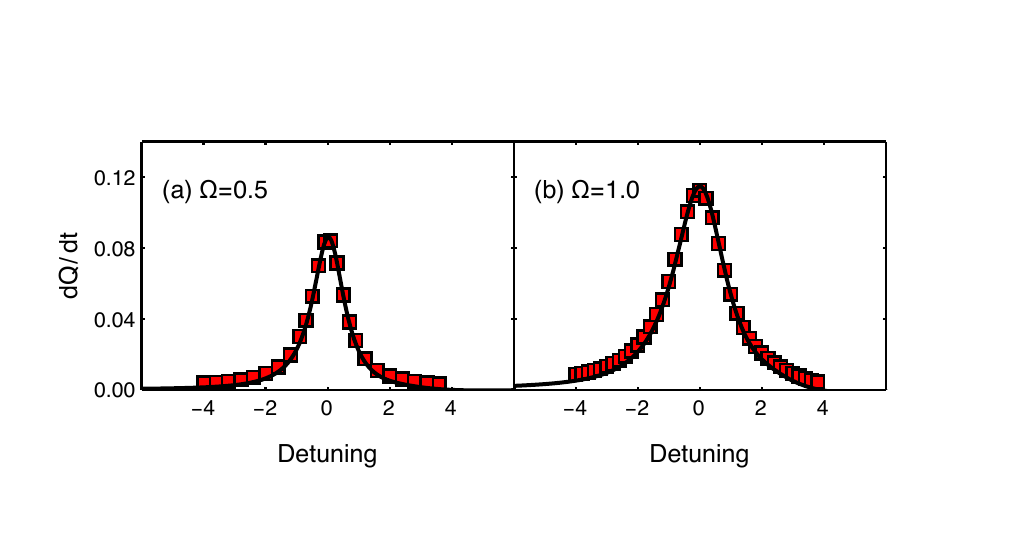}\end{center}
\caption{Heat absorption rates, as a function of the detuning, computed using the full Bloch-Redfield equation (solid black curves) and a numerically exact method (red squares), for two different Rabi frequencies. The Rabi frequencies $\Omega$ are (\textbf{a}) 0.5 and (\textbf{b}) 1.0.\label{figcoolsnumerics}}
\end{figure}

In Fig.\ \ref{figcoolsnumerics} we compare between the cooling spectra
predicted by the full Bloch-Redfield equation with those obtained from the exact
numerical method \cite{popovic_quantum_2021}. The numerical method
simulates the time-evolution of the open quantum
system, using discrete timesteps. For these simulations we have taken
a timestep $dt=0.05$, and computed the heat current, at a time
$t=30.0$, from the difference in the heat transfer at two times. The
mean heat transfer is computed by evaluating the annotated reduced
density matrix, $\rho_{u,S}(t)$, and computing the finite difference
approximation to the derivative in Eq. (\ref{eqmeanheat}) from the
values of $\Tr \rho_{u,S}(t)$ at $u=0.05$ and $u=0.0$. The numerical
accuracy and convergence of these simulations involves two further
parameters: a maximum number of timesteps retained in the influence
functional, $K$, and a cut-off paramater controlling the truncation of
the singular-value decompositions. We take $K=100$, and use a cut-off
of $10^{-7}$.

We see from Fig.\ \ref{figcoolsnumerics} that the Bloch-Redfield
equation is in excellent agreement with the numerical results. The
non-Lorentzian behaviour of the cooling profile on the high-energy
side, predicted by the full Bloch-Redfield and secular equations, is
present. There is a slight overestimate of the cooling power in the
tails of the profiles, which we believe is because the heat current
has not yet reached its steady-state value at those small values of
the cooling power.

\section{Discussion}

Figs.\ \ref{figcools1} and \ref{figcoolsnumerics} suggest that the
full Bloch-Redfield equation gives an accurate account of the cooling
profile, in both the weak- and strong-driving cases. In the
weak-driving case it agrees with the phenomenological theory, which is
well-justified for weak-driving, while in the strong-driving case it
agrees with the secular theory, which is well-justified
there. Furthermore, it agrees with exact numerical results, in the
strong-driving regime where such simulations are possible. Thus the
Bloch-Redfield equation allows a complete treatment of both regimes,
using a single equation. This conclusion is similar to previous
conclusions on the dynamics and thermodynamics of other open quantum
systems, where the Bloch-Redfield equation has similarly been argued
to provide the most accurate description\
\cite{eastham_bath-induced_2016,hartmann_accuracy_2020,purkayastha_out--equilibrium_2016,kilgour_coherence_2018,liu_coherences_2021}. This
is in spite of the possibility that it
produces unphysical density matrices with non-positive eigenvalues. That possibility does not occur in our results, except at very low temperatures where laser cooling would not, in any case, be expected to operate.

In previous works it has been noted that the secular approximation
does not allow for the presence of bath-induced or noise-induced
coherence~\cite{tscherbul_long-lived_2014} in multilevel systems where
near-degenerate levels have different couplings to the
bath~\cite{eastham_bath-induced_2016}. This phenomenon can play an
important role for the heat currents, as has been pointed out
previously for quantum absorption refrigerators
\cite{kilgour_coherence_2018,liu_coherences_2021}. Its significance in
our case can be seen by comparing the secular result (where there is
no bath-induced coherence) and the Bloch-Redfield result (where there is) in
Fig.~\ref{figcools1}. When $\delta=0, \Omega=0$ our Hamiltonian has
two degenerate eigenstates, but the form of those eigenstates depends
on how the limit is taken: for $\delta=0,\Omega\neq 0$ they are
$|\pm\rangle=|g_u\rangle \pm |e\rangle$, but for
$\Omega=0, \delta\neq 0$ they are $|g_u\rangle$ and $|e\rangle$. The
naive form of secular approximation in Eq. (\ref{eq:secmaster})
produces a dissipator which populates the states $|+\rangle$ and
$|-\rangle$, and destroys coherences between them. However, we observe
that the phonon bath couples only to
$|g_u\rangle\propto |+\rangle+|-\rangle$, and not to
$|e\rangle \propto |+\rangle-|-\rangle$, so in the weak-driving case
the correct dissipator should affect the population of the first
combination, while leaving that of the second undamped. This means
that there are undamped coherences in the $|\pm\rangle$ basis, which
survive in the steady-state~\cite{eastham_bath-induced_2016}, and
produce corrections to the heat currents relative to the results of
the secular approximation.

%Authors should discuss the results and how they can be interpreted from the perspective of previous studies and of the working hypotheses. The findings and their implications should be discussed in the broadest context possible. Future research directions may also be highlighted.

%%%%%%%%%%%%%%%%%%%%%%%%%%%%%%%%%%%%%%%%%%
%\section{Conclusions}

%This section is not mandatory, but can be added to the manuscript if the discussion is unusually long or complex.

%%%%%%%%%%%%%%%%%%%%%%%%%%%%%%%%%%%%%%%%%%

%%%%%%%%%%%%%%%%%%%%%%%%%%%%%%%%%%%%%%%%%%
\authorcontributions{Conceptualization, methodology, software, analysis, investigation, C.M., L.T.T., P.E. Writing --- original draft preparation, P.E.; writing---review and editing, C.M., L.T.T.; visualization, supervision, P.E. All authors have read and agreed to the published version of the manuscript.}

%\funding{Please add: ``This research received no external funding'' or ``This research was funded by NAME OF FUNDER grant number XXX.'' and  and ``The APC was funded by XXX''. Check carefully that the details given are accurate and use the standard spelling of funding agency names at \url{https://search.crossref.org/funding}, any errors may affect your future funding.}

\funding{This research was funded by the Irish Research Council grant number GOIPG/2017/1091. Some calculations were performed on the Boyle cluster maintained by the Trinity Centre for High Performance Computing. This cluster was funded through grants from the European Research Council and Science Foundation Ireland.}

\dataavailability{The data generated in this study are available in Zenodo.}
%
%  dx.doi.orgIn this section, please provide details regarding where data supporting reported results can be found, including links to publicly archived datasets analyzed or generated during the study. Please refer to suggested Data Availability Statements in section ``MDPI Research Data Policies'' at \url{https://www.mdpi.com/ethics}. You might choose to exclude this statement if the study did not report any data.} 

\acknowledgments{We thank D. Segal for helpful comments on the
  manuscript, and G. Fux, B. Lovett, and J. Keeling for discussions
  and assistance with the TEMPO and PT-TEMPO codes.}

\conflictsofinterest{The authors declare no conflict of interest.} 

%% Optional
%\sampleavailability{Samples of the compounds ... are available from the authors.}

%%%%%%%%%%%%%%%%%%%%%%%%%%%%%%%%%%%%%%%%%%
%% Only for journal Encyclopedia
%\entrylink{The Link to this entry published on the encyclopedia platform.}

%%%%%%%%%%%%%%%%%%%%%%%%%%%%%%%%%%%%%%%%%%
%% Optional
%\abbreviations{The following abbreviations are used in this manuscript:\\

%\noindent 
%\begin{tabular}{@{}ll}
%MDPI & Multidisciplinary Digital Publishing Institute\\
%DOAJ & Directory of open access journals\\
%TLA & Three letter acronym\\
%LD & Linear dichroism
%\end{tabular}}

\reftitle{References}

% Please provide either the correct journal abbreviation (e.g. according to the “List of Title Word Abbreviations” http://www.issn.org/services/online-services/access-to-the-ltwa/) or the full name of the journal.
% Citations and References in Supplementary files are permitted provided that they also appear in the reference list here. 

%=====================================
% References, variant A: external bibliography
%=====================================
\externalbibliography{yes}
\bibliography{lasercooling.bib}

\end{paracol}
\end{document}